\documentstyle [prl,aps,twocolumn,epsf,psfig]{revtex}
\newskip\humongous \humongous=0pt plus 1000pt minus 1000pt
\def\caja{\mathsurround=0pt}
\def\eqalign#1{\,\vcenter{\openup1\jot \caja
    \ialign{\strut \hfil$\displaystyle{##}$&$
    \displaystyle{{}##}$\hfil\crcr#1\crcr}}\,}
\newif\ifdtup

\relax

\begin{document}

\newcommand{\newc}{\newcommand}

\newc{\be}{\begin{equation}}
\newc{\ee}{\end{equation}}
\newc{\ba}{\begin{eqnarray}}
\newc{\ea}{\end{eqnarray}}
\newc{\bea}{\begin{eqnarray}}
\newc{\eea}{\end{eqnarray}}
\newc{\D}{\partial}
\newc{\ie}{{\it i.e.} }
\newc{\eg}{{\it e.g.} }
\newc{\etc}{{\it etc.} }
\newc{\etal}{{\it et al.}}

\newc{\ra}{\rightarrow}
\newc{\lra}{\leftrightarrow}
\newc{\no}{Nielsen-Olesen }
\newc{\tp}{'t Hooft-Polyakov }
\newc{\lsim}{\buildrel{<}\over{\sim}}
\newc{\gsim}{\buildrel{>}\over{\sim}}
\draft \preprint{DEMO-HEP-99/01 March 99}
\title
{Superconducting Semilocal Stringy (Hopf) Textures\cite{prep}}

\bigskip

\author{L. Perivolaropoulos
\cite{also}}

\address{Institute of Nuclear Physics,
N.C.S.R. Demokritos \\ 153 10 Athens; Greece\\ e-mail:
leandros@mail.demokritos.gr \\ \vspace{5mm}{\it Talk presented at
the NATO Advanced Study Institute of the ESF Network on
'Topological Defects and the Non-Equilibrium Dynamics of Symmetry
Breaking Phase Transitions' at Les Houches, France 16-26/2/1999.}}

\date{\today}
\maketitle

\begin{abstract}

The dynamics of texture-like configurations are briefly reviewed.
Emphasis is given to configurations in 2+1 dimensions which are
constructed numerically. Confirming previous semi-analytical
studies it is shown that they can be stabilized by partial gauging
of the vacuum manifold (semilocality) in a finite range of
parameter space. When these configurations are extended to 3+1
dimensions (stringy textures) it is shown that they can support
persistent currents if a twist (Hopf charge) is introduced in the
scalar field sector. The pressure induced by these persistent
currents is also studied in closed loops. In the context of a
simple model, twist induced pressure is shown to be insufficient
to stabilize the loops against collapse due to tension\footnote{A
large part of this talk is based on recent unpublished work with
T. Tomaras \cite{unpublished}}.

\end{abstract}

\narrowtext


\noindent
\section{Introduction}
A key feature of  texture-like topological defects is the fact
that the topological charge emerges by integrating over the whole
physical space (not just the boundary). These defects have played
an important role in both particle physics and cosmology. Typical
examples are the skyrmion\cite{skyrme} which offers a useful
effective model for the description of the nucleon and the global
texture\cite{turok} where an instability towards collapse of the
scalar field configuration has been used to construct an appealing
mechanism for structure formation in the universe.

A typical feature of this class of scalar field configurations are
instabilities towards  field rescalings which usually lead to
collapse and subsequent decay to the vacuum via a localized highly
energetic event in space-time. The property of collapse is a
general feature of global field configurations in 3+1 dimensions
and was first described by Derrick\cite{derrick}. This feature is
particularly useful in a cosmological setup because it provides a
natural decay mechanism which can prevent the dominance of the
energy density of the universe by texture-like defects. At the
same time, this decay mechanism leads to a high energy event in
space-time that can provide the primordial fluctuations for
structure formation.

In the particle physics context where a topological defect
predicted by a theory can only be observed in accelerator
experiments if it is at least metastable, the above instability is
an unwanted feature. A usual approach to remedy this feature has
been to consider effective models where non-renormalizable higher
powers of scalar field derivatives are put by hand. This has been
the case in QCD where chiral symmetry breaking is often described
using the low energy 'pion dynamics' model. Texture-like
configurations occur here and as Skyrme first pointed out they may
be identified with the nucleons (Skyrmions)\cite{skyrme}. Here
textures are stabilized by non-renormalizable higher derivative
terms in the quantum effective action. However no one has ever
found such higher derivative terms with the right sign to
stabilize the Skyrmion.

An alternative approach to stabilize texture-like configurations
is the introduction of gauge fields\cite{rubakov,bt,bt1} which can
be shown to induce pressure terms in the scalar field Lagrangian
thus balancing the effects of Derrick-type collapse. In the case
of complete gauging of the vacuum manifold however, it is possible
for the texture configuration to relax to the vacuum manifold by a
continuous gauge transformation that can remove all the gradient
energy (the only source of field energy for textures) from the
non-singular texture-like configuration\cite{turok}. This
mechanism of decay via gauge fields is not realized in singular
defects where the topological charge emerges from the boundaries.
In these defects, singularities, where the scalar field is 0, can
not be removed by continuous gauge transformations.

Recent progress in semilocal defects has indicated that physically
interesting models can emerge by a partial gauging of the vacuum
manifold of field theories. This partial gauging (semilocality)
can lead to new classes of stable defect solutions that can
persist as metastable configurations in more realistic models
where the gauging of the vacuum is complete but remains
non-uniform. A typical example is the semilocal string\cite{va91}
whose embedding in the standard electroweak model has led to the
discovery of a class of metastable 2+1 dimensional field
configurations in this model\cite{vp}.

In the context of texture-like configurations, the concept of
semilocality can lead to an interesting mechanism for
stabilization. In fact the semilocal gauge fields are unable to
lead to relaxation of the global field gradient energy because
they can not act on the whole target space. They do however induce
pressure terms in the Lagrangian that tend to resist the collapse
induced by the scalar sector. Therefore they have the features
required for the construction of stable texture-like
configurations in renormalizable models without the adhoc use of
higher powers of derivatives.

The goal of this lecture is to demonstrate the stabilization
induced by semilocal gauge fields in the context of a simple model
that supports stable {\it Superconducting Semilocal Stringy
Texture} (SSST) field configurations. The extension of the form of
the field configuration and its embedding in realistic models is
currently in progress.

The structure of this talk is the following: I will first give a
brief review of the field structure and dynamics of textures in
1+1 and 3+1 dimensions. Then I will focus on the 2+1 dimensional
case and show how can semilocal gauge fields stabilize a
collapsing scalar field configuration. The extension to 3+1
dimensions in the form of {\it a stringy texture} and the
introduction of persistent currents by a topological twist of the
scalar field will also be discussed in section II. The mechanism
by which this current pressure could stabilize a collapsing
stringy texture {\it loop} in 3+1 dimensions will be emphasized.
In section III, these concepts are applied to a simple concrete
model where the SSST configurations are constructed numerically. A
virial theorem demonstrating the stability of the configurations
against collapse in 2+1 dimensions will be verified. Closed loops
of SSST will also be discussed and shown to be unstable towards
collapse in 3+1 dimensions despite the pressure induced by the
persistent currents. This instability is present for all parameter
sectors considered in the context of this simple model.

\section{Texture Fields and their Dynamics}

The simplest texture-like configuration appears in 1+1 dimensions
in a field theory with a two-component real scalar field ${\vec
\Phi}=(\Phi_1, \Phi_2)$ which breaks a global O(2) symmetry and
its dynamics is determined by the potential $V({\vec
\Phi})={\lambda \over 4} ({\vec \Phi}^2 - \eta^2)^2$. The vacuum
manifold $M_0$ of this theory is ${\vec \Phi}^2 = \eta^2$ \ie a
circle $S^1$. It has a non-trivial first homotopy group
$\pi_1(S^1) = Z$. Therefore there are topologically non-trivial
field configurations in 1+1 dimensions such that as one travels
along in space, ${\vec \Phi}$ winds once around $M_0$ (Fig. 1a).
As shown in Fig.1a the magnitude of ${\vec \Phi}$ is close to
$\eta$ everywhere and therefore the total energy of the
configuration may be approximated as
\be
E={1\over 2}\int_{-\infty}^{+\infty} dx \; {\vec \Phi}'^2 \equiv T
\ee
\begin{figure}
\centerline{
\psfig{figure=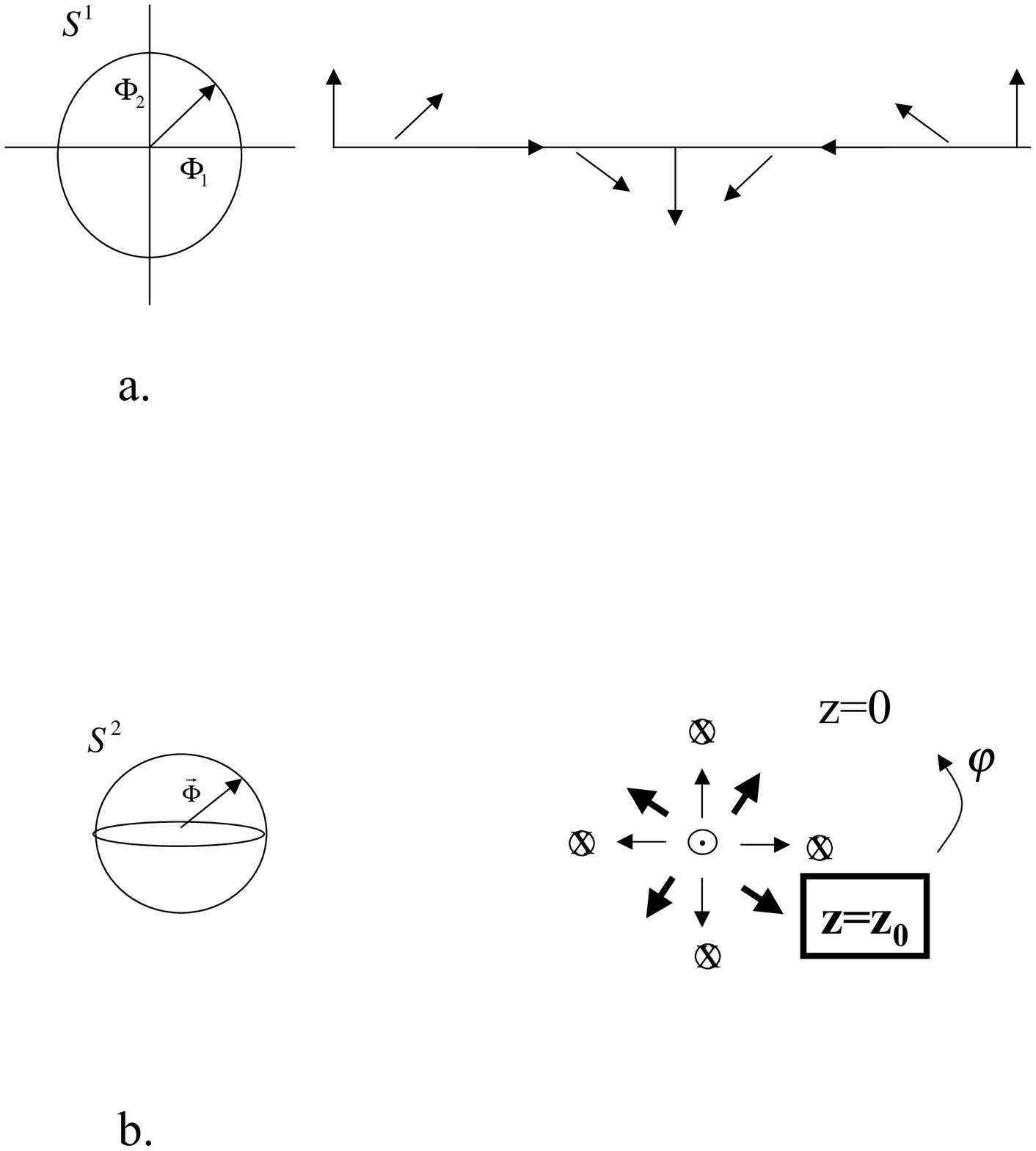,width=3.4in,height=2.6in,angle=0} \psdraft }
Figure 1: (a) The field configuration of a texture in 1+1
dimensions covers completely the vacuum manifold $S^1$ as the
physical space is spanned. (b) In 2+1 dimensions the vacuum
manifold $S^2$ is also covered and there is the possibility of
field twist along the $z$ axis.
\end{figure}

A rescaling of the spatial coordinate $x$ to $\alpha x$ with
$\alpha < 1$ leads to a field configuration ${\vec \Phi}(\alpha
x)$ which has expanded in space relatively to ${\vec \Phi}(x)$.
The total energy of ${\vec \Phi}(\alpha x)$ is clearly $E_\alpha =
\alpha E < E$ and therefore the dynamics will favor expansion of
the configuration ${\vec \Phi}(x)$ to infinity. Thus texture-like
configurations in 1+1 dimensions are unstable to expansion. This
argument does not hold for configurations with appreciable total
potential energy (\eg domain walls). The potential energy scales
as ${1\over \alpha}$ with re-scaling and can therefore prevent the
expansion triggered by the gradient energy.

The 1+1 dimensional texture can be stabilized forming
configurations known as {\it ribbons}\cite{bt2}. This can be
achieved in two ways. The first includes the introduction of a
potential term $V_1({\vec \Phi})=\mu ({\vec \Phi} - {\vec
\eta})^2$ that explicitly breaks the $O(2)$ symmetry and leads to
a rescaled energy $E_\alpha = \alpha T + V_1/\alpha$ which can be
minimized with respect to $\alpha$. The second is implemented by
the introduction of a cutoff $L$ in the one dimensional space
which, given the boundary conditions, is equivalent to
compactifying it. With this cutoff the energy becomes
$E=\int_{-L}^L dx \; {\vec \Phi}'^2 /2$ and $E_\alpha = \alpha \;
\int_{-\alpha L}^{\alpha L} \;dx \;{\vec \Phi}'^2 /2$ which can
also be minimized with respect to $\alpha$.

The texture configuration in 1+1 dimensions can be generalized to
3+1 dimensions by considering a field theory describing a global
symmetry breaking $O(4)\rightarrow O(3)$. Consider for example a
four component scalar field ${\vec
\Phi}=(\Phi_1,\Phi_2,\Phi_3,\Phi_4)$ whose dynamics is described
by the potential \be V({\vec \Phi})={\lambda \over 4} ({\vec
\Phi}^2 - \eta^2)^2 \label{pot31} \ee The initial condition ansatz
\be {\vec \Phi}=(\sin\chi \sin\theta \sin\varphi ,\sin\chi
\sin\theta \cos\varphi,\sin\chi \cos\theta,\cos\chi) \ee with
$\chi (r)$ varying between 0 and $\pi$ as $r$ goes from 0 to
infinity and $\theta$, $\varphi$ spherical polar coordinates,
describes a configuration that winds once around $M_0=S^3$ as the
physical space is covered. The energy of this configuration is of
the form
\be
E=\int_{-\infty}^{+\infty}{1\over 2} ({\vec \nabla}{\vec \Phi})^2
+ V({\vec \Phi})\equiv T + V \ee where we have allowed for
possible small potential energy excitations during time evolution.
A rescaling of the spatial coordinates $r \rightarrow \alpha r$
leads to $E_\alpha = \alpha^{-1} T + \alpha^{-3} V$ which is
monotonic with $\alpha$ and leads to collapse, highly localized
energy and eventual unwinding of the configuration. These highly
energetic and localized {\it events} in spacetime have provided a
physically motivated mechanism for the generation of primordial
fluctuations that gave rise to structure in the
universe\cite{turok}.

The possible stabilization of these collapsing configurations
could lead to a cosmological overabundance and a cosmological
problem similar to the one of monopoles, requiring inflation to be
resolved. At the same time however it could lead to observational
effects in particle physics laboratories. There are at least two
ways to stabilize a collapsing texture in 3+1 dimensions. The
first is well known and includes the introduction of higher powers
of derivative terms in the energy functional. These terms scale
like $\alpha^p$  ($p>0$) with a rescaling and can make the energy
minimization possible thus leading to stable {\it skyrmions}.
Stable {\it Hopfions}\cite{Hopfions} (solitons with non-zero Hopf
topological charge) have also been constructed recently by the
same method. The second method of stabilization is less known (but
see ref. \cite{rubakov,bt}) and can be achieved by introducing
gauge fields that partially cover the vacuum manifold. An example
is the Lagrangian\cite{core}
\begin{equation}
{\cal L} = -{1\over 4} F^a_{\mu \nu}F^{a\mu \nu} + {1\over 2}
D_\mu \Phi^a D^\mu \Phi^a + {1\over 2} (\D_\mu \Phi^4)^2 - V
\label{model2}
\end{equation}
describing the dynamics of an O(3) gauge field $A_{\mu}^a$ coupled
to the three components ($\alpha = 1,2,3$) of the scalar field
${\vec \Phi} = (\Phi_1, \Phi_2, \Phi_3, \Phi_4)$ whose dynamics is
determined by the potential (\ref{pot31}).

The field strength and the covariant derivative are given by
$F^a_{\mu \nu} = \D_\mu A ^a_\nu - \D_\nu A^a_\mu + g
\epsilon^{abc}A^b_\mu A^c_\nu$ and $D_\mu \Phi^a \equiv \D_\mu
\Phi^a + g \epsilon^{abc} A^b_\mu \Phi^c$, respectively.

The above is a simple extension of the Georgi-Glashow O(3) model
with one classically relevant parameter
\be
\tilde\beta = {g\over \sqrt{\lambda}} \ee as revealed after the
rescaling $x^i \to x^i /(v \sqrt{\lambda})$, $\Phi^a \to v
\Phi^a$, $\Phi^4 \to v \Phi^4$, and $A^a_\mu \to v A^a_\mu$.

For $\tilde\beta=0$,  (\ref{model2}) possesses an O(4) global
symmetry. The ansatz describing a semilocal texture in 3+1
dimensions is of the form

\begin{equation}
\eqalign{ {\vec \Phi}&=(f(r) \sin\theta \sin\varphi, f(r)
\sin\theta \cos\varphi,f(r) \cos\theta, G(r)) \cr A^a_i &=
\epsilon_{aij} {x^j \over r} W(r)  \label{texture-ansatz} }
\end{equation}
with $f(r)$ and $W(r)$ necessarily vanishing at the origin $r=0$
and at infinity.

It is convenient to define $K(r) \equiv 1-\tilde\beta r W(r)$, in
which case the field equations for the three unknown functions of
the ansatz take the form:
\begin{equation}
\eqalign{ f'' + {{2f'}\over r} - {{2f}\over r^2} K^2 + (1-f^2 -
G^2) f &= 0 \cr K^{\prime \prime} - {{K(K^2 -1)}\over{r^2}} -
\tilde\beta^2 f^2 K &= 0 \cr G'' + {{2G'}\over r} + (1-f^2 - G^2)
G &= 0 \label{monopole-equations} }
\end{equation}
while the corresponding boundary conditions, dictated by the
finiteness of the energy and the field equations at the origin,
are
\begin{equation}
\eqalign{ f(\infty) &= 0, \;\;\; G(\infty) = 1, \;\;\; K(\infty) =
1 \cr f(0) &= 0, \;\;\; G(0) = -1, \;\;\;\;\; K(0) = 1 \label{bc2}
}
\end{equation}

As discussed in the introduction, a complete gauging of the vacuum
manifold would allow the gauge fields to unwind the topological
charge by a continuous gauge transformation and a decay to the
vacuum. This can not be achieved by the semilocal gauging of
(\ref{model2}). It may also be shown that the gauge fields induce
terms that scale as $\alpha^p$ ($p>0$) in the energy corresponding
to (\ref{model2}) thus preventing the collapse. The detailed
analysis of this model is currently in progress. Another concrete
example of this mechanism in a simple model will be discussed in
section III.

Let us now consider textures in 2+1 dimensions which are also
known as Belavin-Polyakov vortices \cite{bp75}. These can form in
theories with global symmetry breaking leading to a vacuum
manifold $S^2$. Consider for example a field theory describing a
global symmetry breaking $O(3)\rightarrow O(2)$ where the vacuum
manifold $M_0$ is $S^2$. This is achieved by a three component
scalar field ${\vec \Phi}=(\Phi_1, \Phi_2, \Phi_3)$ whose dynamics
is described by the potential (\ref{pot31}). The ansatz
\be
(\Phi_u, \Phi_3)=[|\Phi_u|e^{i\varphi},
\Phi_3(\rho)]=[\sin\chi(\rho)\;e^{i\varphi},\; \cos\chi(\rho)]
\label{anz2} \ee where $\chi(\rho)$ varies between 0 and $\pi$ as
$\rho$ goes from $0$ to infinity (Fig. 1b), describes a
configuration that winds once around $M_0=S^2$ as the physical
space $R^2$ is covered.

Allowing for small potential energy excitations during evolution
leads to an energy of the form
\be
E=\int_{-\infty}^{+\infty}d^2 x \; \; {1\over 2} ({\vec
\nabla}{\vec \Phi})^2 + V({\vec \Phi})\equiv T + V \ee which after
coordinate rescaling becomes $E_\alpha = T + \alpha^{-2} V$ and
the configuration is weakly collapsing if the potential term
becomes significant while otherwise it is neutral with respect to
rescaling. Stabilization can now be achieved in two steps: First
we introduce a potential energy term \be V_1({\vec
\Phi})=\mu(\Phi_3 - \eta)^2 \ee
 which explicitly breaks $O(3)$ and
is therefore non-vanishing at finite distance. Then a rescaling
leads to $E_\alpha = T + \alpha^{-2} V_1$ which implies collapse.
The collapse may now be halted by the introduction of partial
gauging (semilocality) through a $U(1)$ gauge field. As shown in
section III, a rescaling now leads to $E_\alpha = T
-\alpha^{-1}V_2 + \alpha^{-2}V_1$ where $T$, $V_1$ and $V_2$ are
energy components independent of $\alpha$ and therefore $E_\alpha$
can be minimized with respect to $\alpha$ leading to stability and
to a virial theorem relating the terms $V_1$ and $V_2$.

The 2+1 dimensional configuration (\ref{anz2}) may easily be
extended to 3+1 dimensions as a stringy texture by assuming
uniformity along the z-axis (Fig.2). A more interesting case
arises if we allow $\Phi_u$ to twist as we move along the z-axis.
This would generalize the ansatz (\ref{anz2}) to
\be
(\Phi_u, \Phi_3)=[|\Phi_u|e^{i\varphi}\; e^{iu(z)}, \Phi_3(\rho)]
\label{anz3} \ee
\begin{figure}
\centerline{
\psfig{figure=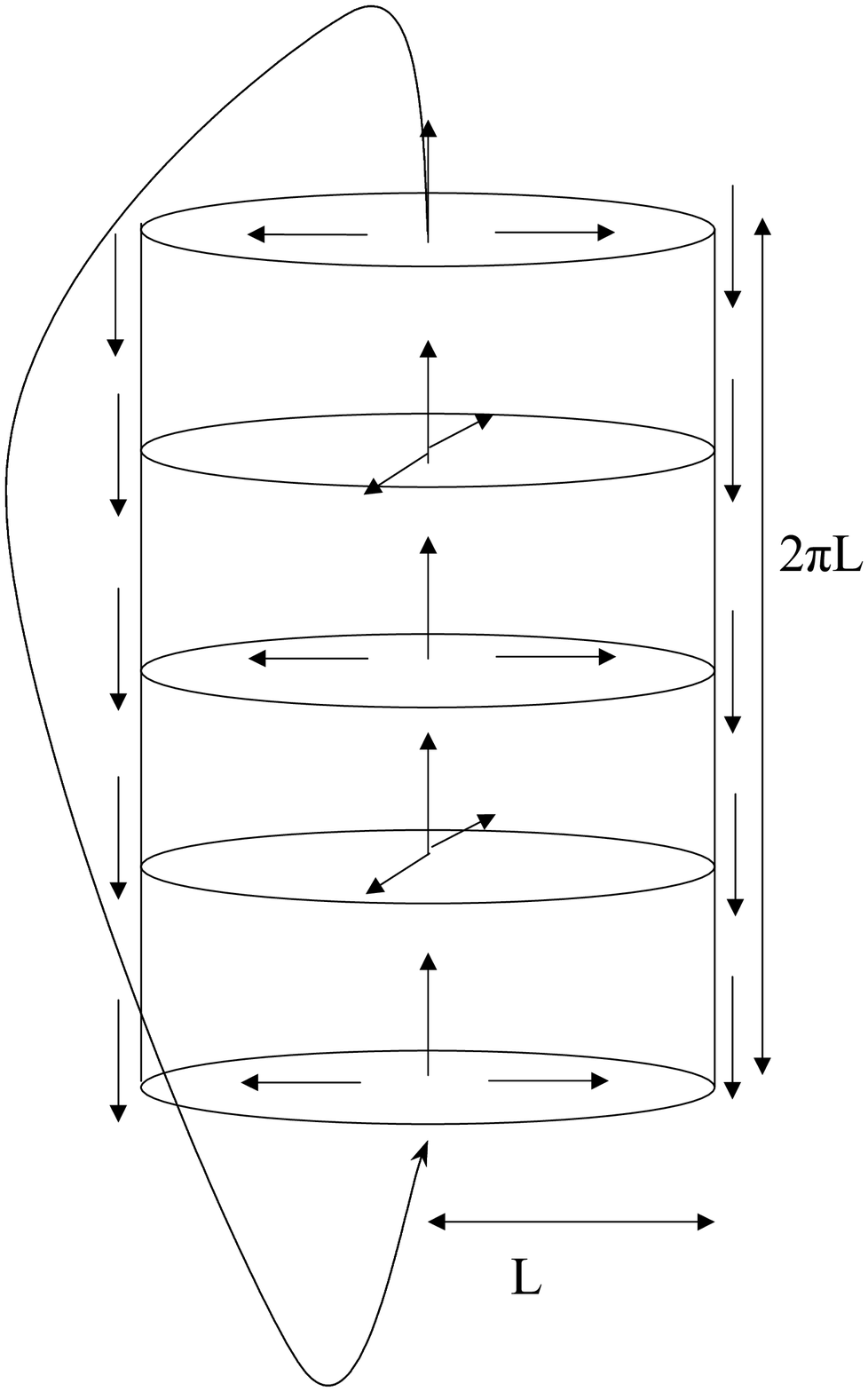,width=3.4in,height=2.6in,angle=0} \psdraft }
Figure 2:  A large loop of stringy texture can be approximated by
a cylinder with periodic boundary conditions.
\end{figure}
The configuration (\ref{anz3}) can also be used to describe a
closed loop of twisted stringy texture in physical space provided
that the following identifications are made.
\begin{enumerate}
\item
The radius $L$ of the cylinder in Fig. 2 is identified with the
radius of the closed loop.
\item
The length of the cylinder is identified with $2\pi L$ \ie the
length of the loop.
\item
The points along a vertical line of the surrounding surface of the
cylinder are identified with the center of the loop.
\item
Points along the vertical line on the opposite side are identified
with infinity.
\end{enumerate}
This approximation of a closed loop by a finite size cylinder has
the advantage of retaining cylindrical symmetry but it fails to
distinguish the center of the loop from infinity. It should
therefore only be trusted for relatively large closed loops
($L>>\eta^{-1}$). The identification of the two faces (upper with
lower) of the cylinder induces a twist topological charge defined
as \be N_t={1\over {2\pi}} \int_0^{2\pi L}\;{{du}\over {dz}}\; dz
\ee in addition to the usual $N_w$ charge of $\pi_2 (S^2)$ of the
two dimensional texture.

The product $N_w N_t$ is the Hopf topological charge $Q_H$. It
appears due to the non-trivial $\pi_3 (S^2)$ which classifies the
twisted loops of stringy textures. The presence of a nonzero Hopf
charge combined with the presence of gauge fields could prevent loop
collapse due to the increase of gradient
energy of the twisted component as the loop collapses (cylinder
shrinks). It will be seen however in section III that this
pressure is not sufficient to balance the tension of the closed
stringy texture in order to stabilize it\footnote{This behavior of
collapse has also been seen in the Hopf
textures\cite{HopfTextures} which differ from the ones discussed
here (section III) in two aspects: First the O(3) symmetry is not
explicitly broken and second gauge fields are not present in the
Hopf textures studied previously}.

\section{Application on a Simple Model}

In order to make the ideas described in the previous sections more
concrete let us consider a simple model\cite{bt,unpublished}
described by the Lagrangian density
\begin{equation}
\eqalign{ &{\cal {L}}=-{1\over 4} F_{\mu \nu} F^{\mu \nu} + {1
\over 2} (D_\mu \Phi)^\dagger D^\mu \Phi +
\partial_\mu \Phi_3 \partial^\mu \Phi_3 \cr
&- V(\Phi, \Phi_3) + {1\over 2} m^2 A_\mu A^\mu \label{model1} }
\end{equation}
where $\Phi =\Phi_1 + i \Phi_2$, $F_{\mu\nu}=\partial_\mu A_\nu-
\partial_\nu A_\mu$,
$D_\mu = \partial_\mu - i e A_\mu$ and

\begin{equation}
V(\Phi, \Phi_3)={\lambda \over 4}(\Phi^\dagger \Phi + \Phi_3^2-
\eta^2)^2 +{k^2 \over 8}(\Phi_3 -\eta)^4 \label{potential1}
\end{equation}

The energy density of the static configuration is
\be
\eqalign{ &{\cal E}={1\over 4} F_{i j} F^{i j} + {1 \over 2} (D_i
\Phi)^\dagger D^i \Phi +
\partial_i \Phi_3 \partial^i \Phi_3 \cr
& - V(\Phi, \Phi_3) + {1\over 2} m^2 A_i {A^i} \label{energy 1} }
\ee where $i,j = 1,2,3$. We now use the following twisted
semilocal stringy texture ansatz
\begin{equation}
\eqalign{ &\Phi = f(\rho) e^{i N_w \varphi}  e^{i u(z)}\, , \;\;
\Phi_3 = g(\rho) \cr &{\vec A} = {\hat e}_\varphi \;a(\rho) +
{\hat e}_z \;b(\rho) \label{string-ansatz} }
\end{equation}
where $0\leq z \leq 2\pi L$. This ansatz describes approximatelly
a large circular loop of SSST in a model with explicitly broken
O(3) symmetry and an additional twist $u(z)$ of the $\Phi$ field
along the length of the loop. This length is approximated by a
large length on the z-axis. The assumption that the radius of the
loop is L (with $L \gg \eta^{- 1}$) imposes the following boundary
conditions on the field function $u(z)$
\be
u(2\pi L) - u(0) = 2\pi N_t \label{boundary1}
\end{equation}
where $N_t$ is the topological charge due to twist. Assuming in
addition that the twist is uniform along the length of the loop we
obtain
\be
{\dot u}\equiv {{du} \over {dz}}={N_t \over L} \label{dotu} \ee
Since $N_t$ is conserved the configuration supports the following
superconducting current densities \ba j_z &=& {{\delta {\cal
L}}\over {\delta A_z}}= -e({\dot u} - e A_z)f^2 \label{jz} \\
j_\varphi &=& {{\delta {\cal L}}\over {\delta A_\varphi}}=
-e({N_w\over \rho} - e A_\varphi)f^2 \label{jz} \ea The
configuration (\ref{string-ansatz}) viewed as a loop in
three-dimensions represents a mapping from space (a large `three
sphere') onto the vacuum manifold (two sphere). The mapping is the
non-trivial Hopf fibration $S^3 \longrightarrow S^2$ with fiber
$S^1$. The corresponding Hopf topological charge is
\be
Q_H = {1\over {8\pi}} \int_{S^2} \epsilon_{abc} \Phi_a d\Phi_b
d\Phi_c = N_w \cdot N_t \label{hopf} \ee With the ansatz
(\ref{string-ansatz}) the energy density (\ref{energy 1}) takes
the form
\be
\eqalign{ &{\cal E} = {1\over 2}(a^\prime + {a \over \rho})^2 +
{1\over 2} {b^\prime}^2 + {1\over 2} f^{\prime  2} + {1\over 2}
({N_w \over \rho} - e a)^2 f^2 \cr &+ {1\over 2} ({\dot u} - e
b)^2 f^2 + {1\over 2} g^{\prime 2} + {\lambda \over 4} (f^2 + g^2
- \eta^2)^2 \cr &+ {k^2 \over 8} (g-\eta)^4 + {1\over 2} m^2 (a^2
+ b^2) \label{energy2} } \ee We now impose the following rescaling
\be
[field] \longrightarrow [field]\, {m\over \sqrt{2\lambda}} \; , \;
{\vec x} \longrightarrow {{\vec x} \over m} \ee where [field]
refers to the field functions $f(\rho),\; g(\rho),\; a(\rho)$ and
$b(\rho)$ while ${\vec x}$ refers to $\rho$ and $z$. The energy
density ${\cal E}$ becomes
\be
\eqalign{ &{\cal E} = {m^4 \over {2 \lambda}} [ {1\over
2}(a^\prime + {a \over \rho})^2 + {1\over 2} b^{\prime 2} +
{1\over 2} f^{\prime 2} + {1\over 2} ({N_w \over \rho} - {\tilde
e} a)^2 f^2 \cr & + {1\over 2} ({\dot u} - {\tilde e} b)^2 f^2 +
{1\over 2} g^{\prime 2} + {\lambda \over 4} (f^2 + g^2 - m_H^2)^2
\cr &+ {{\tilde k}^2 \over 8} (g-m_H)^4 + {1\over 2} (a^2 + b^2) ]
\label{energy3} } \ee where ${\tilde e} = e/\sqrt{2\lambda}$,
${\tilde k}=k/\sqrt{2\lambda}$ and $m_H = \sqrt{2\lambda} \; \eta
/m$.

By extremizing the static energy density (\ref{energy3}) we obtain
the field equations
\begin{equation}
\eqalign{ &-(a' + {a \over \rho})' - {\tilde e} ({n\over \rho}
-{\tilde e} a)f^2 + a =0 \cr &-{1\over \rho} (\rho b')' - {\tilde
e} ({N_t \over L} - {\tilde e} b)f^2 + b = 0 \cr &-{1\over \rho}
(\rho f)' - ({N_w \over \rho} - {\tilde e} a)f +({N_t \over L} -
{\tilde e} b)^2 f \cr &+ {1\over 2} (f^2 + g^2 - m_H^2) f = 0 \cr
&-{1\over \rho} (\rho g)' + {1\over 2} (f^2 + g^2 - m_H^2) f +
{{\tilde k}^2 \over 2} (g-m_H)^3 = 0 \label{fieldequations} }
\end{equation}
The boundary conditions used for the solution of the system
(\ref{fieldequations}) at the center of the SSST ($\rho=0$) and at
the center of the loop ($\rho = L$) are dictated by the finiteness
of the energy and the field equations at the origin and may be
written as
\begin{equation}
\eqalign{ &f(0)=0\, , \;\; a(0)=0 \, , \;\; \rho b'(0)=0 \, , \;\;
\rho f'(0)=0\cr &\rho g'(L)=0\, , \;\; g(L)=\eta \, , \;\; b(L)=0
\, , \;\; {\vec B}(L)\equiv {\vec \nabla}\times {\vec A}(L) = 0
\label{boundary2} }
\end{equation}
We have used a relaxation method with locally variable mesh size
to solve this system of equations in order to identify parameter
sectors where solutions exist. The validity of the derived
solutions was verified by checking that they satisfy to a good
approximation virial conditions that can be obtained analytically
from the energy functional (\ref{energy3}) by appropriate
rescalings. In particular by demanding that the solution is an
extremum of the total energy in two dimensions and therefore its
energy does not change to first order by a rescaling $\rho
\rightarrow \alpha \; \rho$ we have the condition ${{\delta
E}\over {\delta \alpha}} = 0$ which implies
\be
v_1 \equiv {{E_1 - E_2}\over {E_1 + E_2}} = 0 \label{v1} \ee where
\be
\eqalign{ E_1 &= \int_0^L d\rho \; N_w \; {\tilde e} \; f^2 \; a
\cr E_2 &= {1\over 2} \int_0 ^L d \rho \;  \rho \;[a^2 (1+{\tilde
e}^2 f^2)+{1\over 4} (f^2 + g^2 - m_H^2)^2 \cr &+ {{\tilde k}^2
\over 4} (g - m_H)^4 + b^2 + ({{N_t}\over L} - {\tilde e}\; b)^2
f^2] \label{e1e2} } \ee An other virial condition may be obtained
by rescaling the gauge field $a(\rho)$ whose boundary conditions
are insensitive to rescaling ($a(0) = a(L) = 0$). By rescaling $a
\rightarrow \alpha \; a$ and demanding ${{\delta E}\over {\delta
\alpha}} = 0$ we obtain
\be
v_2 \equiv {{E_1 - E_3}\over {E_1 + E_3}} = 0 \label{v2} \ee where
\be
E_3 =\int_0^L d\rho \; \rho \; [(a' + {a\over \rho})^2 + (1+
{\tilde e}^2 f^2)\;  a^2 ] \label{e3} \ee All the solutions obtained
by the relaxation method satisfied the above virial conditions to
a very good approximation. In all cases we had $v_1 \lsim
10^{-3}$ and  $v_2 \lsim 10^{-3}$.

As mentioned above, the model considered here reduces to that of
Ref. \cite{bt} for zero twist ($N_t = 0$). Thus we can test
numerically the semi-analytic approximate results of Ref.
\cite{bt} by finding parameter sectors where stable solutions
exist in the limit $N_t \rightarrow 0$. In this limit the only
relevant parameters for the existence and stabily of solutions
were shown in Ref. \cite{bt} to be \ba a&\equiv& {{\tilde k}^2
\over {\tilde e}^2}\\ b&\equiv& {2 \over {{\tilde e}^2 m_H^4}} \ea
\begin{figure}
\centerline{ \psfig{figure=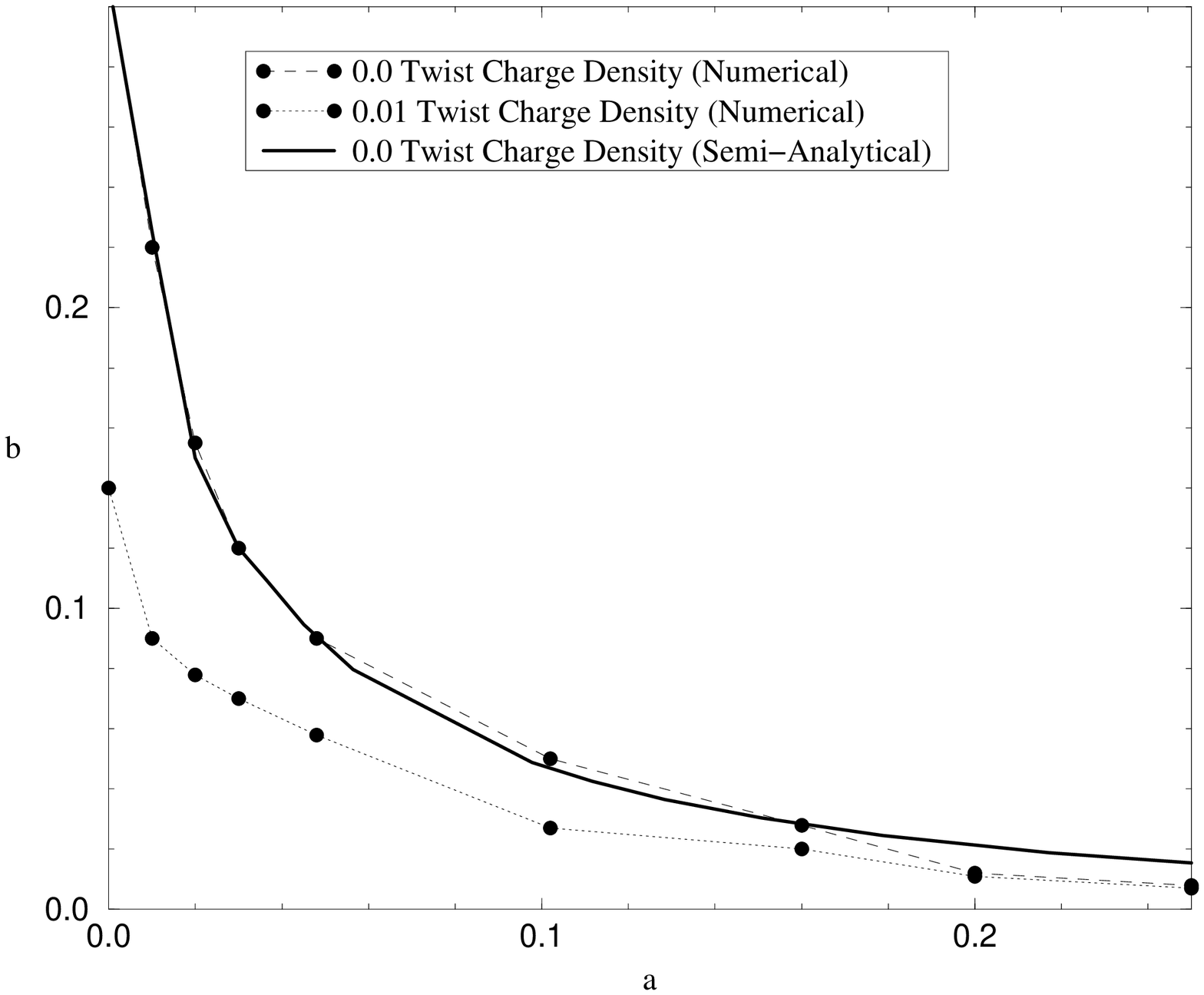,height=3.2in,angle=-90}
\psdraft } Figure 3:  For values of the parameters a and b below
the dashed ($N_t/L = 0$) and dotted ($N_t / L = 0.01$) lines we
have found numerically, classically stable solutions. No solutions
were found above these lines. The dashed line is in very good
agreement with the approximate semi-analytical results of Ref.
\cite{bt} obtained for $N_t =0$.
\end{figure}
We have first solved numerically the coupled system of field
equations (\ref{fieldequations}) for zero twist and found the
range of parameters in a-b space where solutions exist. The
results are shown in Fig. 3 (dashed line) and their agreement with the
approximate results of Ref. \cite{bt} is very good. We then
introduce a small twist ($N_t =10$ with $L=1000$) and solve the
field equations again for the same parameter values. The parameter
region for solution existence gets reduced as expected and is also
shown in Fig. 3 (dotted line) in the a-b plane. The derived solutions have also
been varied numerically using a large number of smooth random
fluctuations and we verified that the perturbations always tend to
increase the total energy. Thus the derived solutions which are
obviously extrema of the total energy correspond to minima (not
maxima) and are stable solutions of the field equations.

In order to examine the effectiveness of pressure induced by the
superconducting currents we have looked for solutions in the
parameter sectors corresponding to high current density (twist
charge density) just before the solution is lost due to increased
energy density. We have found that even though there are solutions
with relatively high twist charge density (provided that $m_H$ is
also large enough) the effect of the pressure is not nearly enough
to stabilise the shrinking loop configuration and produce a total
energy local minimum for a finite value of L. Thus even though the
solutions obtained are local minima of the energy as two
dimensional configurations, when considered in three dimensions as
loops they are expected to shrink monotonically towards zero loop
radius ($L=0$) due to tension.

\begin{figure}
\centerline{ \psfig{figure=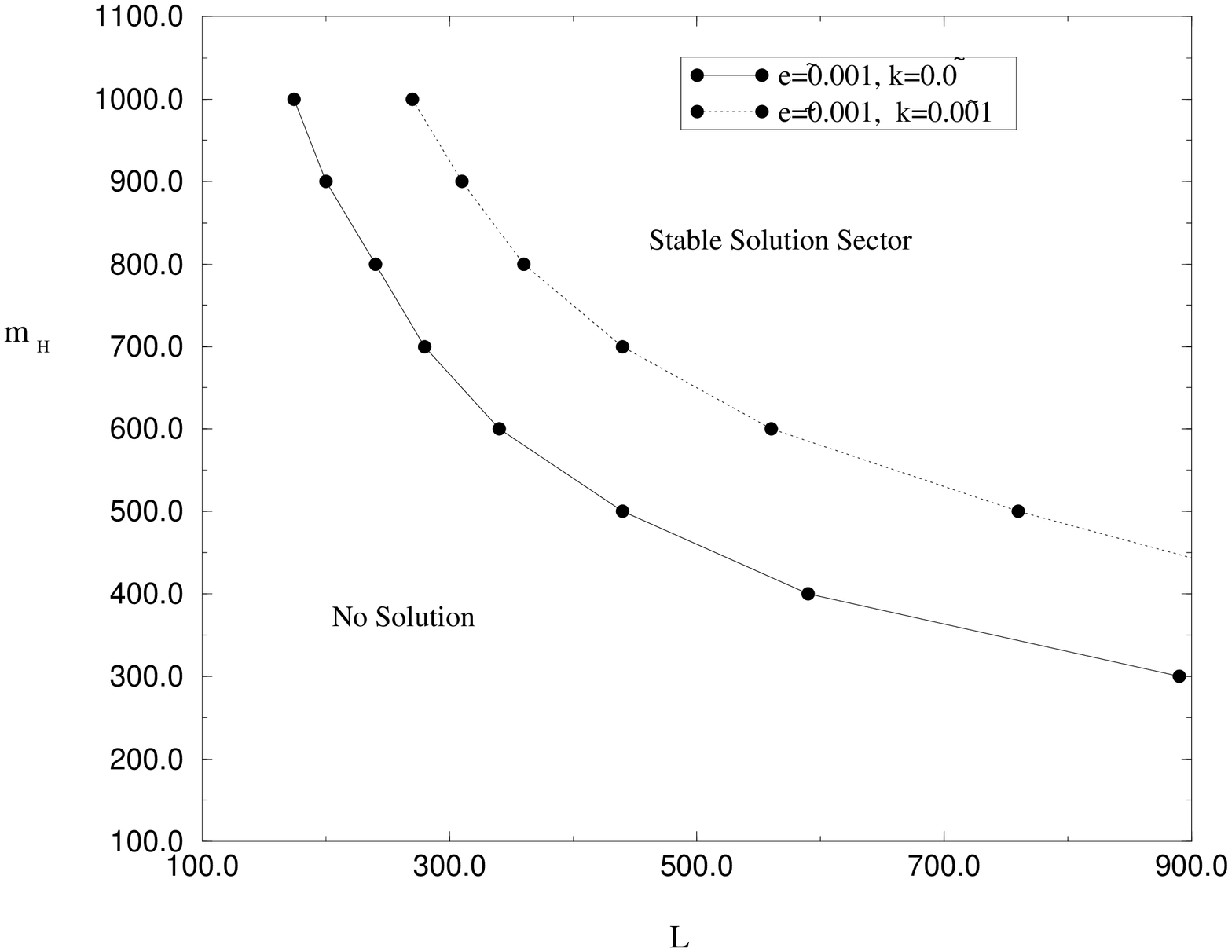,height=3.2in,angle=-90}
\psdraft } Figure 4:  The parameter sectors where stable solutions
were found in the parameter region of high $N_t$. In all cases we
had $N_t = 70$ and $N_w = 1$. The parameter space was scaned by
reducing $L$ for each fixed $m_H$ ($300\leq m_H \leq 1000$).
\end{figure}

The parameter sectors where highly twisted stable solutions have
been found are shown in Fig. 4 as a function of the loop radius
$L$. Notice that the increase of $m_H$ improves the stability of
the solution and allows the loop to shrink further before the
solution disappears due to very large twist topological charge.
Thus for larger $m_H$ the length $L$ of the loop can drop further
before the instability sector is reached. We anticipate that this
effect is similar to the current quenching which is seen in the
case of the usual superconducting strings \cite{witten}. As the
loop shrinks due to tension the twist topological charge density
increases and at a critical density it becomes energetically
favorable for the field $f$ to become 0 in order to unwind part of
the twist topological charge and therefore reduce the
superconducting current. This effect is known as {\it current
quenching}. The critical size of the loop for current quenching is
expected to decrease as the parameter $m_H$ increases because it
becomes more costly energetically for $f$ to vanish. This behavior
is seen in Fig. 4. At quenching the solution breaks down because
in our approximation of uniform large loop we have not allowed
dynamics in ${\dot u}(z)$ assuming that the twist charge density
remains uniform at all times. This approximation is valid for
currents smaller than the quenching currents but in order to
describe the actual quenching process when the conservation of
$N_t$ is violated one needs a full 3-dimensional simulation of the
loop evolution\cite{vs94}. This is outside the scope of the
present study.

\begin{figure}
\centerline{
\psfig{figure=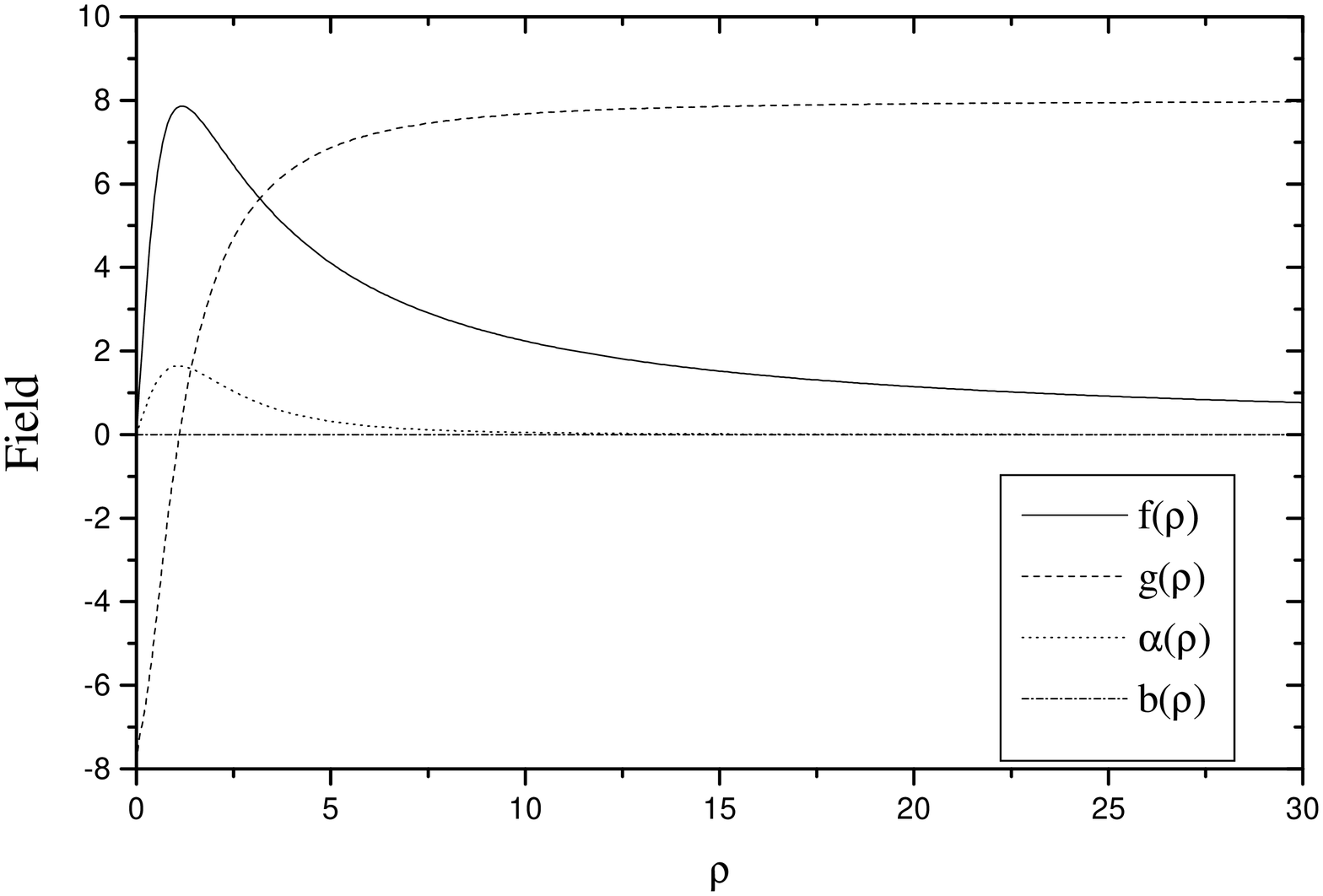,width=3.4in,height=2.6in,angle=0} \psdraft }
Figure 5:  The field profile for $L=950$. The rest of the
parameters were: $m_H = 8$, ${\tilde k}=0.01$, ${\tilde e} = 0.1$,
$N_t = 1$ and $N_w = 1$. The virial theorems were satisfied ($v_1
\simeq 6\times 10^{-4} \; \; v_2\simeq 10^{-4}$
\end{figure}

In Fig. 5 we show the field functions $f(\rho)$ and $g(\rho)$ for
some choice of parameters.

An interesting question that needs to be addressed in the context
of the model studied is the question of spring formation
\cite{witten,chht88}. The total energy of the large loop
configuration ($L \gg \eta^{-1}$) may be written as:
\be
E(L) =2\pi \int_0^{2\pi L} dz \int_0^L d\rho \; \rho \; {\cal
E}(f,g,a,b) \label{totenergy} \ee where ${\cal E}$ in its rescaled
form is given by equation (\ref{energy3}). Given the presence of
an $L$-dependent term (the twist gradient)  in ${\cal E}$ it could
have been anticipated that $E(L)$ has a minimum at some
$L=L_{spring}$. This would imply the formation of a loop
stabilized by the pressure induced by the twist charge which acts
against the tension of the loop. These objects were anticipated by
analytical arguments in the Witten models\cite{witten,chht88} of
superconducting strings but no such solution has been found
explicitly so far.

Clearly, spring formation could occur only for $L$ large enough so
that the solution exists \ie $L_{spring} > L_{quench}$. If this
minimum of the energy could be achieved at some $L_{spring} >
L_{quench}$ then at the $L_{quench} (m_H)$ shown in Fig. 3, $E(L)$
would have a negative derivative with respect to L i.e. the total
energy $E$ would tend to decrease towards its minimum as $L$
increased from $L_{quench}$ towards $L_{spring}$. We have checked
all points at $L_{quench}$ shown in Fig. 3 and found that ${{dE}
\over {dL}}|_{L=L_{quench}}
> 0$. Therefore we conclude that for all the parameter sectors
we investigated no spring solutions exist.

\section{Conclusion -Outlook}
The main points of this talk can be summarized as follows
\begin{itemize}
\item
Semilocality can stabilize textures in 2+1 dimensions.
\item
Twisted Superconducting Semilocal Stringy Texture (SSST)
configurations exist for a finite sector of parameter space.
\item
No SSST loops stabilized by current pressure (springy textures)
were found in the simple model considered.
\end{itemize}

An alternative way to stabilize SSST loops is the introduction of
angular momentum whose conservation can stabilize loops against
collapse more effectively than twist pressure. Loops stabilized by
angular momentum are known as vortons\cite{vortons}. Potential
extensions of this work include the study of the effects of
angular momentum and the embedding of these configurations in
realistic two-Higgs electroweak models\cite{bt1}.

\section{Acknowledgements}
This work is the result of a network supported by the European Science
Foundation.
The European Science Foundation acts as catalyst for the development of
science by bringing together leading scientists and funding agencies to
debate, plan and implement pan-European initiatives.
This work was also supported by the EU grant CHRX-CT94-0621 as well
as by the Greek General Secretariat of Research and Technology grant
$\Pi$ENE$\Delta$95-1759.

\vfill

\eject

\end{document}